\documentclass{article}

\usepackage{arxiv}

\usepackage[utf8]{inputenc} 
\usepackage[T1]{fontenc}    
\usepackage{hyperref}       
\usepackage{url}            
\usepackage{booktabs}       
\usepackage{amsfonts}       
\usepackage{nicefrac}       
\usepackage{microtype}      
\usepackage{lipsum}
\usepackage{graphicx}
\graphicspath{ {./images/} }

\title{Cost-Effective Deep Learning Infrastructure with NVIDIA GPU}

\author{
  Aatiz Ghimire \\
   Central Department of Physics\\
   Tribhuvan University\\
   Kirtipur, Kathmandu 44613, Nepal\\
  \texttt{hello@aatizghimire.com.np} \\
   \And
 Shahnawaz Alam \\
  Herald College Kathmandu\\
  University of Wolverhampton\\
  Naxal, Kathmandu 44600, Nepal \\
  \texttt{shahnawaz.alam@heraldcollege.edu.np} \\
  \And
 Siman Giri \\
  Herald College Kathmandu\\
  University of Wolverhampton\\
  Naxal, Kathmandu 44600, Nepal \\
  \texttt{siman.giri@heraldcollege.edu.np} \\
\And
Madhav Prasad Ghimire \\
   Central Department of Physics\\
   Tribhuvan University\\
   Kirtipur, Kathmandu 44613, Nepal\\
  \texttt{madhav.ghimire@cdp.tu.edu.np} \\
  }

\begin{document}
\maketitle
\begin{abstract}
The growing demand for computational power is driven by advancements in deep learning,
the increasing need for big data processing, and the requirements of scientific simulations for academic
and research purposes. Developing countries like Nepal often struggle with the resources needed to
invest in new and better hardware for these purposes. However, optimizing and building on existing
technology can still meet these computing demands effectively. To address these needs, we built a
cluster using four NVIDIA GeForce GTX 1650 GPUs. The cluster consists of four nodes: one master
node that controls and manages the entire cluster, and three compute nodes dedicated to processing
tasks. The master node is equipped with all necessary software for package management, resource
scheduling, and deployment, such as Anaconda and Slurm. In addition, a Network File Storage
(NFS) system was integrated to provide the additional storage required by the cluster. Given that the
cluster is accessible via ssh by a public domain address, which poses significant cybersecurity risks,
we implemented fail2ban to mitigate brute force attacks and enhance security. Despite the continuous
challenges encountered during the design and implementation process, this project demonstrates how
powerful computational clusters can be built to handle resource-intensive tasks in various demanding
fields.
\end{abstract}

\keywords{Deep Learning Infrastructure \and Beowulf Cluster \and High-Performance Computing
(HPC) \and GPU Cluster Architecture}

\section{Introduction}
Deep learning has become a cornerstone of modern artificial intelligence, driving advances in various fields such as natural language processing, computer vision, and autonomous systems. However, training deep learning models often requires substantial computational resources, which can be prohibitively expensive for many institutions and researchers. Cloud-based GPU solutions, while offering scalability, come with recurring costs and potential data privacy concerns, making them less feasible for long-term or resource-intensive projects. \cite{10.1145/3447545.3451184} \cite{lawrence2017comparing} \cite{munhoz_performance_2023}

High-performance computing (HPC) or Clusters is increasingly shifting towards heterogeneous GPU-based systems, driven by the growing demand for massively parallel computing in deep learning and scientific applications\cite{han_quantitative_2019}. With the exponential growth of AI-driven research and simulations, traditional CPU-based architectures often struggle to keep up with computational demands. GPUs, with their high throughput and parallel processing capabilities, have emerged as a crucial component in accelerating tasks such as model training, data analysis, and large-scale simulations.

Building an in-house, cost-effective deep learning infrastructure offers a practical alternative to address these challenges. Local clusters equipped with GPUs not only reduce operational costs but also provide greater control over data and computational resources. However, achieving a balance between performance, scalability, and affordability remains a significant challenge, especially for setups that use consumer-grade GPUs and standard hardware components. \cite{vaihela2022performance}

This study focuses on designing and implementing a cost-effective deep learning cluster using readily available hardware and open source software. The objectives are to evaluate the feasibility of such a system, address its limitations, and highlight its potential for supporting computational research in resource-constrained environments. Using locally available resources and optimizing configurations, this study aims to contribute to the growing demand for accessible and efficient AI infrastructure.

The rest of this paper is structured as follows: Section 2 provides an overview of related studies. Section 3 walks through the step-by-step process of building the GPU cluster, including the essential software and its role at each stage. In Section 4, we discuss the results, while Section 5 presents a summarized configuration and software stack. Finally, Section 6 concludes the paper with key takeaways.

\section{Related Works}
\label{sec:relatedworks}

The Beowulf Raspberry Pi cluster has gained attention as an affordable and scalable alternative for parallel computing applications. Various studies have explored its implementation, highlighting its potential for educational and experimental purposes \cite{raspberry_2020, cicirello_design_2024, vargas-perez_designing_2022, moses_mwasaga_implementing_2020, mollova_laboratory_2018, penyala_raspberry_2020}. While Raspberry Pi clusters provide a cost-effective way to introduce students and researchers to distributed computing concepts, their performance is limited due to the low computational power of individual nodes. Furthermore, Raspberry Pi lacks dedicated GPU acceleration, restricting its usability for deep learning and large-scale parallel computing tasks. This limitation makes it unsuitable for workloads requiring significant matrix operations, such as AI model training.

Another notable low-cost alternative is the Odroid-XU4 board, which offers more processing capability than Raspberry Pi. Studies have demonstrated its viability in HPC education and small-scale computing clusters \cite{alvarez_teaching_2018}. Odroid-XU4 features an ARM-based architecture with multi-core processing, making it more efficient for parallel computations than Raspberry Pi. However, similar to Raspberry Pi, it lacks dedicated GPU support, making it impractical for deep learning tasks that heavily rely on GPU acceleration. While software optimization can help mitigate some performance constraints, these devices are not suitable for handling large-scale AI and scientific computing workloads that demand high floating-point performance.

Recycling old computers by integrating them into a Rocks cluster is another cost-efficient strategy for setting up HPC environments \cite{rao_analysis_2020}. Many academic institutions have repurposed outdated lab machines to build functional compute clusters, extending their usability for scientific computing and parallel processing workloads \cite{kumar_design_2018}. However, these older systems often rely solely on CPUs and lack modern GPUs, which are essential for accelerating deep learning computations. This absence of GPU support significantly limits their effectiveness in AI research and computational tasks that require tensor operations. While CPU-based clustering can still be useful for general-purpose parallel processing, it falls short in performance compared to modern GPU-accelerated systems.

Despite the availability of low-cost solutions, pre-compiled DGX OS, designed specifically for deep learning systems, does not support commodity hardware, restricting its deployment to specialized NVIDIA DGX systems \cite{noauthor_release_nodate}. The NVIDIA DGX systems, optimized for deep learning workloads, has been widely adopted in AI research due to its integrated software stack and high-performance GPUs \cite{emil_creation_nodate, majee_deepops_2024, noauthor_deploying_nodate, noauthor_deploying_2021}. Unlike the previously mentioned hardware solutions, NVIDIA DGX systems is built with dedicated NVIDIA GPUs, enabling high-speed parallel computations essential for deep learning. However, its high cost makes it inaccessible for many institutions and researchers, pushing them to explore alternative, budget-friendly solutions that attempt to bridge the gap between affordability and computational efficiency.

\section{MATERIALS AND METHODS}
\label{sec:materialsandmethods}

On the hardware side, our aim was to maintain a minimal configuration to make use of commonly available computers in standard computing labs. Our setup consisted of four computers equipped with GPUs, connected via a switch using CAT6 Ethernet cables for network communication.
We assigned the four computers as Master, C1, C2, and C3 nodes.

\begin{figure}[h]
    \centering
    \includegraphics[width=0.5\textwidth]{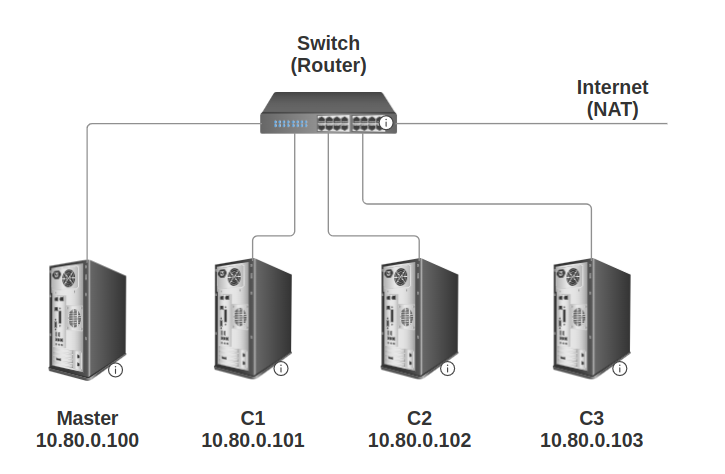}
    \caption{Cluster Network Configuration}
    \label{fig:example}
\end{figure}

The basic configuration of our cluster is shown in the figure above. The computers are connected to a switch using Ethernet cables. Alternatively, a router can be used instead of a switch. One port of the switch is connected to the internet, enabling remote access to the cluster. We configured Network Address Translation (NAT) to convert the cluster's local private IP addresses (Master node IP) into a global public IP address for internet accessibility.

Initially, we attempted to set up the cluster using Ubuntu Server 22.04.1. However, it required extensive package installations, had higher configuration complexity, and lacked enterprise-level features such as robust security, package stability, and infrequent updates. To address these limitations, we switched to Rocky Linux 9.4, which provides RHEL features. We performed a clean installation of Rocky Linux on each cluster node and updated them to the latest version. For simplicity in installation and cluster configuration, we only set up the root user with same password and did not create additional user accounts at this stage. \cite{rockylinuxInstallingRocky}

The next step was to configure static IP addresses, which are essential for inter-server communication needed for tasks such as SLURM, MPI, and data transfer. To achieve this, we used "nmtui", a Network Manager tool, to assign static IP addresses to the nodes, as illustrated in the figure above.

To streamline communication within the cluster, we configured the \textit{'/etc/hostname'} file on all nodes. The host file maps IP addresses to hostnames, allowing us to use simple names instead of IP addresses for communication. We assigned \textit{'10.80.0.100'} to the master node and sequential IP addresses (e.g., \textit{'10.80.0.101', '10.80.0.102', '10.80.0.103'}) to the compute nodes 'c1', 'c2', and 'c3'. Additionally, we updated the '\textit{/etc/hostname}' file on each node to define its hostname, ensuring consistent identification across the cluster. This configuration simplifies node communication and is essential for seamless cluster operations, including MPI, SLURM, and other distributed tasks.

 
We installed the SSH package on all nodes to enable secure communication within the cluster. To streamline operations, we configured passwordless SSH, allowing quick and seamless login between nodes without requiring repeated password entry. This setup simplifies administrative tasks and facilitates efficient execution of distributed processes across the cluster. \cite{tecmintSetupPasswordless}

All logins to the cluster are routed through the master node, which serves as the central point for user access and management. The master node is equipped with all necessary software for package management, resource scheduling, and deployment, ensuring efficient coordination and operation of the cluster. This centralized setup simplifies user management and resource allocation across the cluster.

\begin{figure}[h]
    \centering
    \includegraphics[width=0.5\textwidth]{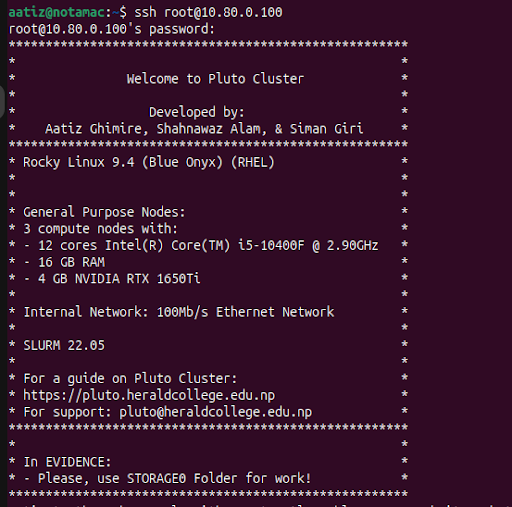}
    \caption{Cluster Login Screen via SSH}
    \label{fig:example}
\end{figure}

To streamline software deployment across multiple nodes in the cluster, we utilize pdsh, a parallel shell tool that enables simultaneous execution of commands. First, pdsh is installed on the head node using \textit{dnf install -y pdsh}, ensuring SSH access is configured for passwordless authentication across all compute nodes. With pdsh, software packages can be installed in parallel, significantly reducing deployment time. For instance, executing \textit{pdsh -w node[1-4] 'sudo dnf install -y package-name'} installs the required software on all designated nodes simultaneously. This method enhances administrative efficiency, ensuring consistency across the cluster without manual intervention, making it well-suited for high-performance computing environments running Rocky Linux.\cite{rittmanmeadLinuxCluster}


Additionally, maintaining time synchronization across the cluster is crucial for job scheduling, logging accuracy, and distributed computing. We achieve this by installing and configuring Network Time Protocol (NTP) on all nodes. Using pdsh, NTP can be installed in parallel with \textit{pdsh -w node[1-4] 'sudo dnf install -y chrony'}. The NTP service is then enabled and started with \textit{pdsh -w node[1-4] 'sudo systemctl enable --now chronyd'}. The head node is configured as the primary NTP server, while compute nodes sync their time with it. This setup ensures clock consistency, preventing synchronization issues that could disrupt HPC workloads and SLURM job execution.

Similary, we installed MUNGE, which is a lightweight authentication service essential for secure, token-based communication between nodes in clusters. It simplifies user authentication, integrates seamlessly with job schedulers like SLURM, and provides scalability and ease of maintenance across clusters, making it ideal for efficient cluster management. \cite{githubGitHubSergioMEVslurmfordummies}

We added an additional 1TB of storage to the system to accommodate large-scale dataset downloads and enable long-term storage of results. The storage was mounted at \textit{`/mnt/storage0`} and bind-mounted to each user's directory as \textit{`/home/username/storage0`}. This setup allows users to access the additional storage directly from their home directory, and we recommend utilizing this location for data management. \cite{rackspaceBindMounts}

We set up a Network File System (NFS) on \textit{'/mnt/storage0'} to enable efficient file sharing between the server and client nodes in our cluster. The configuration involved installing the \textit{`nfs-utils`} package, creating shared directories with appropriate permissions, and defining NFS exports in the \textit{`/etc/exports`} file. We configured the firewall to allow NFS services and created mount points on the client nodes. The NFS shares were mounted manually for testing and configured in \textit{`/etc/fstab`} for persistence after reboots. This setup ensures secure and reliable file sharing across the cluster. \cite{howtoforgeMountRocky}

We utilized FreeIPA for centralized user management and synchronization across all cluster nodes. FreeIPA provided seamless integration of user authentication and access control, ensuring consistent user accounts and credentials throughout the system. This streamlined administrative tasks, enhanced security, and simplified user management in the cluster environment.

\begin{figure}[h]
    \centering
    \includegraphics[width=0.5\textwidth]{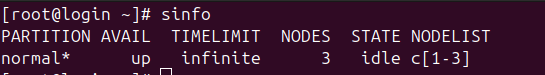}
    \caption{Slurm Command sinfo for listing cluster}
    \label{fig:example}
\end{figure}

User management in HPC clusters requires a balance between security and ease of administration. While NIS is widely used in many HPC environments due to its simplicity, it is considered less secure than modern alternatives \cite{barone_designing_2022, kumar_distributed_2023}. FreeIPA, which integrates LDAP and Kerberos, offers enhanced security and centralized authentication but is more complex to install and configure \cite{arisal_managing_2019}. Our cluster utilizes FreeIPA with LDAP to provide a robust identity management system, ensuring secure user authentication and streamlined access control. However, for clusters prioritizing ease of use over security, NIS remains a simpler alternative despite its vulnerabilities.

Slurm is an open-source workload manager designed for efficient resource allocation, job scheduling, and queue management in high-performance computing (HPC) clusters. To set up Slurm in our cluster, we installed the necessary packages on all nodes and configured the `slurm.conf` file to match our cluster's specifications. The configuration file was distributed to all nodes, and appropriate permissions were set for Slurm directories. On the controller node, the Slurm controller daemon (`slurmctld`) was enabled and started, while the Slurm node daemon (`slurmd`) was activated on all worker nodes. The installation was verified using commands like `sinfo` for cluster information and `srun` for test jobs. This setup ensures efficient job scheduling and resource management, optimizing the performance of the cluster. \cite{githubGitHubSergioMEVslurmfordummies}.
We configured the master node to support computations but chose not to use it for this purpose. Utilizing the master node for processing tasks would add significant load to it, as it is already responsible for managing multiple tasks, including job scheduling, resource allocation, and acting as the login node for multiple users. Prioritizing these critical functions ensures the stability and efficiency of the cluster.

At the time of setup, NVIDIA driver installation was not supported directly through the Rocky Linux package manager (dnf). Therefore, we manually installed the driver using the `.run` file obtained from the NVIDIA website. The GPU driver was installed first, followed by the installation of CUDA to enable GPU-accelerated computations in the cluster.

To enable GPU integration in SLURM, we configured the `slurm.conf` file to recognize GPU resources by adding the `Gres` (Generic Resources) parameter. Each node was configured with its specific GPU details, including type and count. SLURM's `gres.conf` file was updated accordingly, ensuring seamless allocation of GPU resources to jobs. This setup allowed efficient scheduling and utilization of GPUs across the cluster. \cite{schedmdSlurmWorkload}

We installed Lmod, a Lua-based environment modules system, to efficiently manage software environments in the cluster. Lmod enables users to dynamically load, unload, and switch between different versions of software and their dependencies. Using Lmod, we created module files for various versions of Python, PyTorch, etc., allowing users to easily switch between them based on their project requirements. After installation, we configured the module files directory and integrated Lmod with the shell environment by updating the system's profile files. This setup, powered by Lua scripting for custom module configurations, ensures flexibility and simplicity in managing diverse software stacks, optimizing the cluster for various computational workflows. \cite{lmodTransitionLmod}

We configured Lmod and Lua on all cluster nodes to ensure consistent environment management and seamless job execution across the cluster. Lmod and Lua were installed on each node, and module files were synchronized using a shared directory accessible to all nodes. The shell environment on each node was updated to load Lmod and set the appropriate `MODULEPATH`. This setup ensures that all nodes can access the same software environments, allowing SLURM to execute jobs reliably and efficiently with the required configurations.

\begin{figure}[h]
    \centering
    \includegraphics[width=1\textwidth]{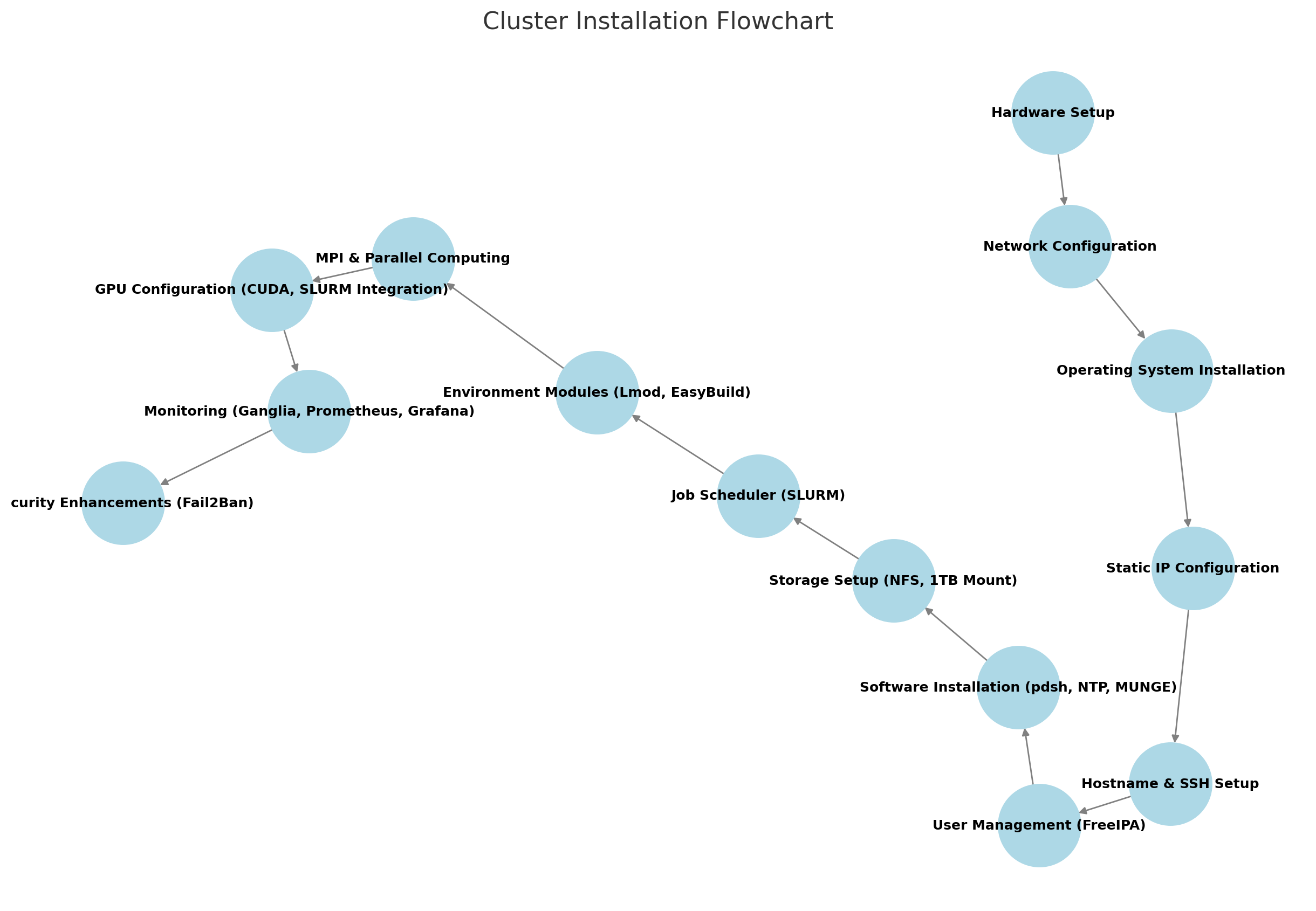}
    \caption{Deep Learning Cluster Installation Flowchart}
    \label{fig:example}
\end{figure}

To further streamline software management, we integrated EasyBuild, an automated framework for building and managing software installations in HPC environments. EasyBuild simplifies the process of compiling and installing complex scientific software by handling dependencies and environment configurations automatically. By using EasyBuild alongside Lmod, we generated module files dynamically, ensuring that all installed software could be easily managed within the cluster's module system. This approach reduces manual configuration efforts while maintaining reproducibility and consistency across all nodes. Additionally, EasyBuild allows us to install optimized versions of libraries and applications, improving performance for computational workloads. \cite{nesuswstutorialsbddlModulesLMod}

While our cluster primarily utilizes EasyBuild, Spack \cite{gamblin2015spack} is also a powerful and flexible alternative for software deployment. Spack enables users to install multiple software versions and manage dependencies efficiently, offering a more modular approach than traditional package managers. Its extensive package repository and customizable builds make it a preferred choice for many HPC environments. Although we opted for EasyBuild due to its structured installation and seamless module integration, Spack remains an excellent choice for clusters requiring greater flexibility in software management.

We installed Anaconda in a shared directory accessible across the cluster and integrated it with Lmod for dynamic environment management. A custom Lmod module file was created to configure Anaconda, setting environment variables like \textit{`PATH`} for seamless activation. Using the shared directory, all cluster nodes can access the Anaconda installation, enabling users to load and unload Anaconda dynamically through Lmod. This centralized setup allows users to create isolated Conda environments, which can be loaded as part of SLURM job scripts, ensuring consistent software environments for diverse computational tasks across the cluster.

We installed multiple versions of Python alongside Anaconda in a shared directory to provide flexibility for various computational tasks. Each Python version was configured with its own Lmod module file, allowing users to dynamically load the desired version as needed. These Python modules, like Anaconda, are accessible across all cluster nodes via the shared directory. Users can load specific Python versions using Lmod commands and utilize them in SLURM job scripts, ensuring compatibility with diverse project requirements while maintaining a consistent and centralized environment management system.

We set up the Message Passing Interface (MPI) using MPICH on the cluster to enable efficient parallel computing. Open MPI was installed on all nodes, ensuring the availability of runtime libraries, compiler wrappers, and development tools. Environment variables such as \textit{'PATH'} and \textit{'LD\_LIBRARY\_PATH'} were configured globally or managed dynamically using Lmod. We verified the installation by compiling and running test MPI programs using 'mpicc' and 'mpirun'. Static IPs and passwordless SSH were configured already before to facilitate seamless communication between nodes. Additionally, MPI was integrated with SLURM, utilizing 'srun' for job execution to leverage SLURM's resource management capabilities. This setup provides a robust framework for scalable parallel computing in the cluster.


The NVIDIA GTX 1650 GPUs in our cluster support CUDA for GPU-accelerated parallel computations but lack the features required for clustering GPUs across nodes, such as GPUDirect RDMA. This limitation prevents the GPUs from being used together in a unified multi-GPU setup. \cite{olmos_advanced_2020} Instead, each GPU operates independently, and tasks must be run on individual GPUs within their respective nodes. While this restricts the scalability of GPU workloads, it remains suitable for single-node GPU computations and smaller parallel tasks.

\begin{figure}[h]
    \centering
    \includegraphics[width=0.7\textwidth]{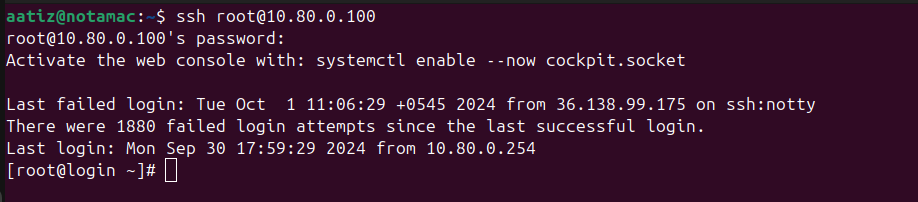}
    \caption{Failed Login attempts in cluster}
    \label{fig:example}
\end{figure}

Monitoring is a critical aspect of HPC cluster management, ensuring system health, performance optimization, and fault detection. We installed Ganglia due to its ease of setup and widespread use in HPC environments, allowing efficient tracking of system metrics across all compute nodes \cite{rao_analysis_2020, mollova_laboratory_2018}. Ganglia provides real-time resource utilization insights, helping to maintain workload balance and optimize job scheduling.

Additionally, we tested Prometheus (on \textit{http://10.80.0.100:9090}) and Grafana (on \textit{http://10.80.0.100:3000}) as alternative monitoring solutions. Prometheus, with its time-series monitoring and alerting capabilities, allows detailed system metric collection via exporters \cite{majee_deepops_2024}. Grafana, when integrated with Prometheus, provides an intuitive visualization interface for tracking CPU, memory, and GPU usage in real time. While both Prometheus and Grafana are viable options under the given IP configurations, we preferred Ganglia for its simpler deployment and efficient monitoring of our SLURM-managed cluster.

Our cluster faced a significant cybersecurity threat daily due to internet connectivity with nearly 2000-6000 failed login attempts to the root account within 24 hours, which we suspect were primarily bot-driven attacks. To mitigate this risk, we disable login with root account in SSH and we installed the Fail2Ban package to enhance SSH security. Fail2Ban monitors login attempts and, after six consecutive failed password entries, automatically blacklists the offending IP address for an hour, blocking further SSH communication. This proactive approach effectively reduced unauthorized access attempts and strengthened the cluster's overall security.

\section{RESULTS AND DISCUSSION}
\label{sec:resultanddiscussion}
The implementation of our cluster setup demonstrates a cost-effective and practical solution for high-performance computing tasks. The master node efficiently managed resource allocation and user authentication, ensuring seamless operation across the cluster. Security was significantly enhanced with the integration of Fail2Ban, which successfully reduced unauthorized login attempts by blacklisting suspicious IP addresses.

The individual GPU performance was validated for CUDA computations, proving effective for single-node tasks despite the hardware limitation of not being able to cluster GPUs. MPI communication over Ethernet was reliable, providing acceptable latency and bandwidth for medium-scale workloads. Additionally, the integration of MPI with SLURM ensured efficient resource scheduling for distributed computations.

The adoption of Lmod and Lua simplified software environment management, enabling users to switch dynamically between various versions of Python, and other required software. The setup of Anaconda further streamlined Python-based workflows, offering flexibility and efficiency for diverse computational tasks. These features significantly improved the user experience in managing software environments within the cluster.

\begin{figure}[h]
    \centering
    \includegraphics[width=0.5\textwidth]{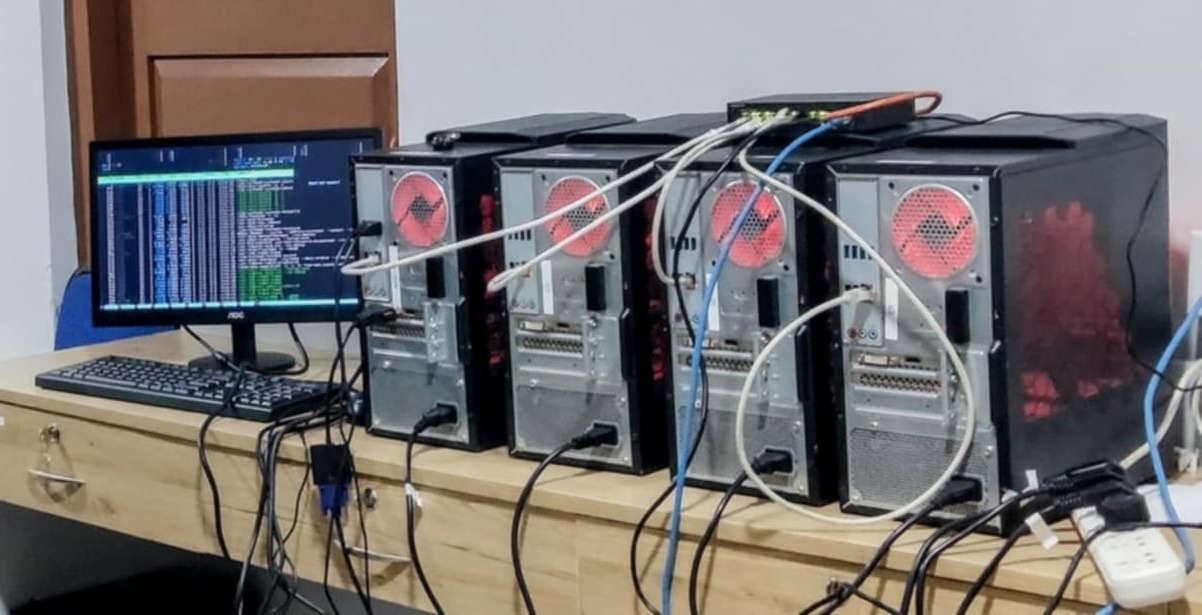}
    \caption{Cluster Setup}
    \label{fig:example}
\end{figure}

The addition of 1TB of shared storage, bind-mounted to user directories, facilitated the handling of large-scale datasets and ensured easy access to results. This enhanced the system's ability to support data-intensive applications and long-term storage needs.

Containerization plays a crucial role in deep learning workflows by providing isolated and reproducible environments. In our cluster, we evaluated both Singularity and Docker as potential solutions for containerized deep learning environments. While Docker is widely used for its flexibility and ease of deployment, it requires root privileges, making it less suitable for shared HPC environments. Conversely, Singularity enables users to run containers securely without requiring elevated permissions, making it a more HPC-friendly alternative. Both containerization tools facilitate seamless execution of deep learning frameworks like TensorFlow and PyTorch, ensuring dependency consistency across different nodes. Based on our evaluation, Singularity is preferred for our cluster due to its security model and compatibility with HPC resource managers like SLURM, but Docker remains a viable option for standalone deep learning projects and development environments.
\cite{emil_creation_nodate} \cite{noauthor_deploying_2021} \cite{noauthor_how_2019}

In summary, the results demonstrate the feasibility of building a robust HPC cluster using commonly available hardware and open-source software. While the system effectively balances cost, scalability, and performance, limitations such as the inability to cluster GPUs and reliance on Ethernet for MPI communication introduce constraints for high-demand, large-scale applications. Future improvements could focus on integrating high-speed interconnects and upgrading to HPC-oriented GPUs to enhance the cluster's performance and scalability further.
\\
\section{TABLES AND COMPARISON}
\label{sec:tables}
The following contains the summary of components and software used:
\begin{table}[!ht]
    \centering
    \begin{tabular}{|l|l|}
    \hline
        Component & Specification \\ \hline
        Master Node & 10.80.0.100, Intel i5 CPU, 16 GB RAM \\ \hline
        Compute Nodes & c1, c2, c3 (10.80.0.101-103), Intel i5 CPU, 16 GB RAM \\ \hline
        GPUs & NVIDIA GTX 1650 (on all nodes) \\ \hline
        Storage & 1TB (shared across all nodes) \\ \hline
        Network & Ethernet (1 Gbps) \\ \hline
    \end{tabular}
\caption{Cluster Configuration Summary}
\end{table}

\begin{table}[!ht]
    \centering
    \begin{tabular}{|l|l|l|}
    \hline
        Software & Version & Purpose \\ \hline
        Rocky Linux & 9.4 & Operating System \\ \hline
        SLURM & 22.05 & Job Scheduling \\ \hline
        Lmod & Installed & Environment Management \\ \hline
        Anaconda & 2023.09 & Python-based Workflows \\ \hline
    \end{tabular}
    \caption{Software Stack}
\end{table}

\section{CONCLUSION}
\label{sec:Conculsion}
Setting up a local GPU cluster within a small department or university presents a highly cost-effective alternative to renting cloud GPUs. Our cost analysis demonstrates that operating a locally installed cluster with four NVIDIA GTX 1650 GPUs (4 x 4GB = 16GB total) incurs a monthly electricity cost of only NPR 5,760 (based on an 800W power consumption per node) in Nepal. In contrast, renting an NVIDIA T4 GPU (16GB) in the cloud costs approximately NPR 123.50 per hour, amounting to NPR 29,640 per month for similar usage \cite{globalpetrolpricesNepalElectricity, googleNVIDIATesla}.

This stark difference underscores the long-term financial benefits of local GPU clusters, particularly in regions like Nepal, where electricity costs are relatively low. Beyond cost savings, a locally managed cluster offers greater control over data privacy, independence from internet connectivity, and uninterrupted access to computational resources, making it an ideal solution for research and education. While the initial hardware investment may seem substantial, many institutions can leverage existing computing labs to set up these clusters efficiently. Ultimately, the autonomy, sustainability, and reduced recurring costs of an in-house GPU cluster make it a far more viable and strategic investment for academic and research institutions compared to reliance on cloud-based alternatives.

\section{ACKNOWLEDGEMENTS}
\label{sec:acknowledgements}
We express our sincere gratitude to Herald College Kathmandu for providing the research space, hardware, and infrastructure essential for this project.

For detailed instructions on installing, and configuring, we have made our setup process and related scripts available at \href{https://github.com/aatizghimire/pluto-cluster}{https://github.com/aatizghimire/pluto-cluster}.
For running this cluster, We have also made slurm with gpu sbatch file and instruction at \href{https://github.com/aatizghimire/deep-learning-gpu-slurm-template}{https://github.com/aatizghimire/deep-learning-gpu-slurm-template}.

\bibliographystyle{unsrt}

\end{document}